\documentclass[sigconf]{acmart}
\AtBeginDocument{%
}
\usepackage{forest}
\usepackage{bbm}

\setcopyright{acmlicensed}
\copyrightyear{2018}
\acmYear{2018}
\acmDOI{XXXXXXX.XXXXXXX}
\acmConference[Conference acronym 'XX]{Make sure to enter the correct
conference title from your rights confirmation email}{June 03--05,
2018}{Woodstock, NY}
\acmISBN{978-1-4503-XXXX-X/2018/06}

\begin{document}

\title{Interpreting Dolphin Vocal Sequences via Multiple Sequence Alignment}

\author{Daniel Kohlsdorf, Denise Herzing}
\email{dkohlsdorf@gmail.com, dlherzing@wilddolphinproject.org }
\affiliation{%
\institution{Wild Dolphin Project}
\city{West Palm Beach}
\state{Florida}
\country{USA}
}
\author{Thad Starner}
\email{thad@cc.gatech.edu}
\affiliation{%
\institution{Georgia Institute of Technology}
\city{Atlanta}
\state{Georgia}
\country{USA}
}

\begin{abstract}
Dolphin communication understanding is essential for uncovering the linguistic complexity and social structures of wild pods. We adapt the ClustalW bio-informatics algorithm to analyze continuous acoustic data, treating vocalizations as high-dimensional spectral feature vectors. By replacing discrete scoring with a continuous Gaussian kernel similarity measure, our framework generates Multiple Sequence Alignment (MSA) visualizations that reveal shared structural patterns. These alignments highlight temporal motifs—such as synchronized burst pulses in aggressive contexts—that are difficult to detect through standard spectrogram inspection. 
\end{abstract}



\begin{teaserfigure}
\includegraphics[width=\textwidth]{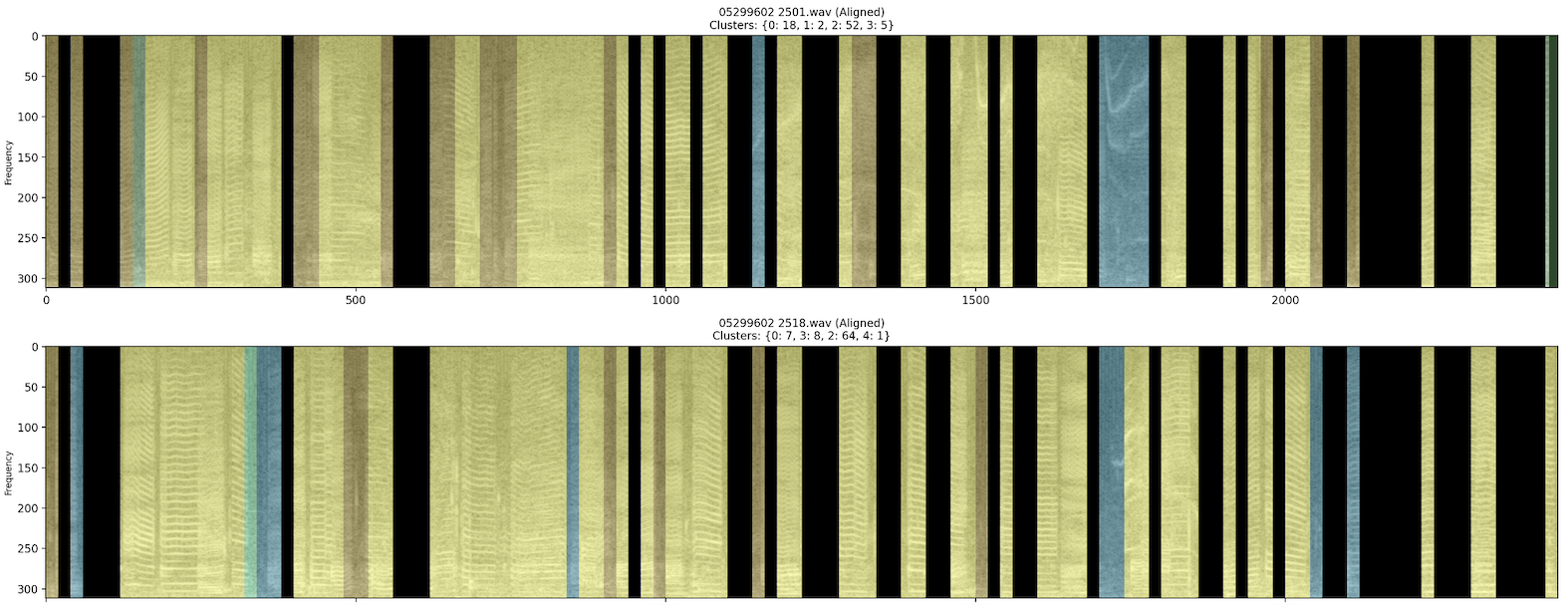}
\caption{A pairwise alignment of two dolphin vocalisation sequences recorded during aggressive behavior.}
\label{fig:teaser}
\end{teaserfigure}

\received{20 February 2007}
\received[revised]{12 March 2009}
\received[accepted]{5 June 2009}

\maketitle

\section{Introduction}
Marine mammalogists investigate the social structures and cognitive capacities of wild dolphins, a field increasingly supported by machine learning. A primary application involves using algorithms to detect recurring patterns within vast vocalization databases. By identifying latent structures in dolphin communication, researchers aim to uncover evidence of linguistic complexity. Furthermore, we believe that a deeper understanding of animal communication can lead to a better understanding of their social structure and enable future application in communicating with animals directly using their own signals. 
%To preserve natural behaviors, we explicitly exclude the artificial playback of any acoustic patterns associated with aggression or social disruption, limiting our active methodology to non-invasive observation
Such communication could more effectively support conservation and monitoring efforts.

Marine mammalogists at the Wild Dolphin Project collect underwater audio and video data from dolphin pods in various behavior contexts in the wild. The data set consists of recordings of spotted dolphins (Stenella frontalis) observed in the Bahamas and spans more than 30 summers. Behaviors include play behavior, nursing, foraging, and aggression, among many others. For example, we know that the sound profile of aggressive behavior is characterized by synchronizing burst pulses (several stacked lines in the spectrogram, see Figure \ref{fig:teaser}). 

In order to highlight structure like the repeated nature of the aggression sound profile, we draw inspiration from bio-informatics, applying sequence alignment techniques to discover structural organization within dolphin acoustic data. In the following, we introduce the concepts of pairwise and multiple sequence alignments, demonstrating how the resulting visualizations can assist marine mammalogists in their search for structure. Furthermore, we illustrate how to adapt the bioinformatics algorithm ClustalW for recordings of dolphin vocalizations. Finally, we present initial results highlighting structural patterns identified across several behavioral contexts, such as aggression.

\section{Ethical Considerations}
While the prospect of direct communication with wild dolphins offers significant benefits for conservation—such as providing a means to warn pods of environmental hazards or anthropogenic noise—it also presents substantial ethical risks. Using synthetic versions of their own acoustic signals could inadvertently disrupt complex social structures, cause psychological stress, or create a dependency on human interaction that alters natural foraging and mating behaviors. Our research prioritizes a non-invasive, "listener-first" approach to mitigate these risks, focusing on decoding signals to better understand dolphin agency and social requirements without imposing human-centric agendas. We recognize that any future attempt at bidirectional communication must be governed by strict protocols that prioritize the dolphins' autonomy and the preservation of their wild behavioral integrity. 

In this paper, our analysis focuses on aggression. Our ethical framework specifically excludes the playback of agonistic or aggressive acoustic patterns, such as synchronized burst pulses. We posit that reproducing these signals carries an unacceptable risk of triggering intra-pod conflict or causing long-term social displacement. By identifying these patterns through machine learning, our goal is strictly to improve human recognition of dolphin social states to avoid researcher interference, rather than to manipulate those states through artificial signaling.

\section{Related Work}

The automated analysis and interpretation of marine mammal vocalizations sits at the intersection of bioacoustics, sequence analysis, and machine learning. Dolphins utilize a sophisticated acoustic repertoire consisting of broadband echolocation clicks, frequency-modulated whistles, and burst-pulsed sounds for navigation, hunting, and social interactions \cite{Janik2018}. However, decoding these communication patterns is historically hindered by manual data extraction challenges and intense ambient underwater noise \cite{Janik2018}. 

To scale analysis, modern machine learning (ML) algorithms and Deep Neural Networks (DNN)  are increasingly deployed to classify dolphin vocalizations \cite{Mueller2021}. For instance, DNNs have successfully categorized distinct whistle types in bottlenose dolphins, linking acoustic signatures to foraging, mating, or socializing behaviors \cite{Yano2020}. These models can detect emergent behaviors or vocal sequences that signal pod coordination and social bonds \cite{Herman2017}. Additionally, ML clustering has been vital for identifying individual callers via unique acoustic signatures \cite{Jensen2024} and tracking the contextual usage of ``signature whistles'' that function as individual identifiers  (i.e., ``names'') across complex social matrices \cite{Janik1998, Sayigh2017, sayigh2023, jensen2024automatic}.

Beyond decoding communication structures, ML models support conservation by modeling anthropogenic noise impacts \cite{Weilgart2007, Simpson2022} and enabling continuous, non-intrusive population monitoring via hydrophones and autonomous drones \cite{Sullivan2021}.

To extract deeper insights from these growing datasets, bioacoustic research has shifted toward modeling the sequential properties of communication. In any learned feature space, we can construct an acoustic model of dolphin communication that explicitly accounts for temporal structure. Our approach leverages Hidden Markov Models (HMMs) to describe how dolphin sounds evolve over time \cite{kohlsdorf_phd, kohlsdorf_icassp, Esfahanian2014, Adi2008, kohlsdorf_interspeech}. This framework mathematically clusters vocalizations in a principled, probabilistic manner, allowing recurring sound patterns to be automatically discovered from raw audio streams \cite{kohlsdorf_phd, kohlsdorf_icassp, Esfahanian2014, Adi2008, kohlsdorf_interspeech}. To mitigate environmental distortion, we train an unsupervised denoising autoencoder to reconstruct clean vocalizations from synthetically corrupted inputs \cite{kohlsdorf_ijcnn}. The decoder mirrors the encoder structure using deconvolution, upsampling, and an LSTM stack to successfully preserve long-range sequence dependencies \cite{kohlsdorf_ijcnn}.

Crucially, linguistic structure can be discovered directly from sequence alignments \cite{abl, kohlsdorf_phd}. Drawing inspiration from computational biology, the ClustalW progressive multiple sequence alignment algorithm has been adapted for dolphin communication analysis \cite{thompson1994clustal}. By treating acoustic feature sequences as biological strings, alignment-based frameworks reveal conserved motifs and structural syntax across variable communicative exchanges. Our proposed method builds upon these sequence alignment and generative paradigms to provide a scalable framework for uncovering the latent syntactic organization of cetacean communication.

Beyond hand-engineered features, self-supervised audio models such as wav2vec 2.0~\cite{wav2vec2} and HuBERT~\cite{hubert} learn general-purpose representations without labeled data, offering an
alternative to Whisper's supervised pretraining that may reduce bias toward human speech. Cross-species bioacoustic foundation models such as BirdNET~\cite{birdnet}, pursue similar goals of general acoustic representation and motif discovery across species, paralleling our own approach. We note that the specific choice of encoder is orthogonal to our core contribution; any of these feature extractors could serve as a drop-in replacement for our fine-tuned, dolphin-specific Whisper model within our alignment framework, which focuses on adapting
sequence alignment itself to continuous acoustic embeddings.

\section{Working with Audio Data}
\begin{figure}[ht]
\centering
\includegraphics[width=0.9 \linewidth]{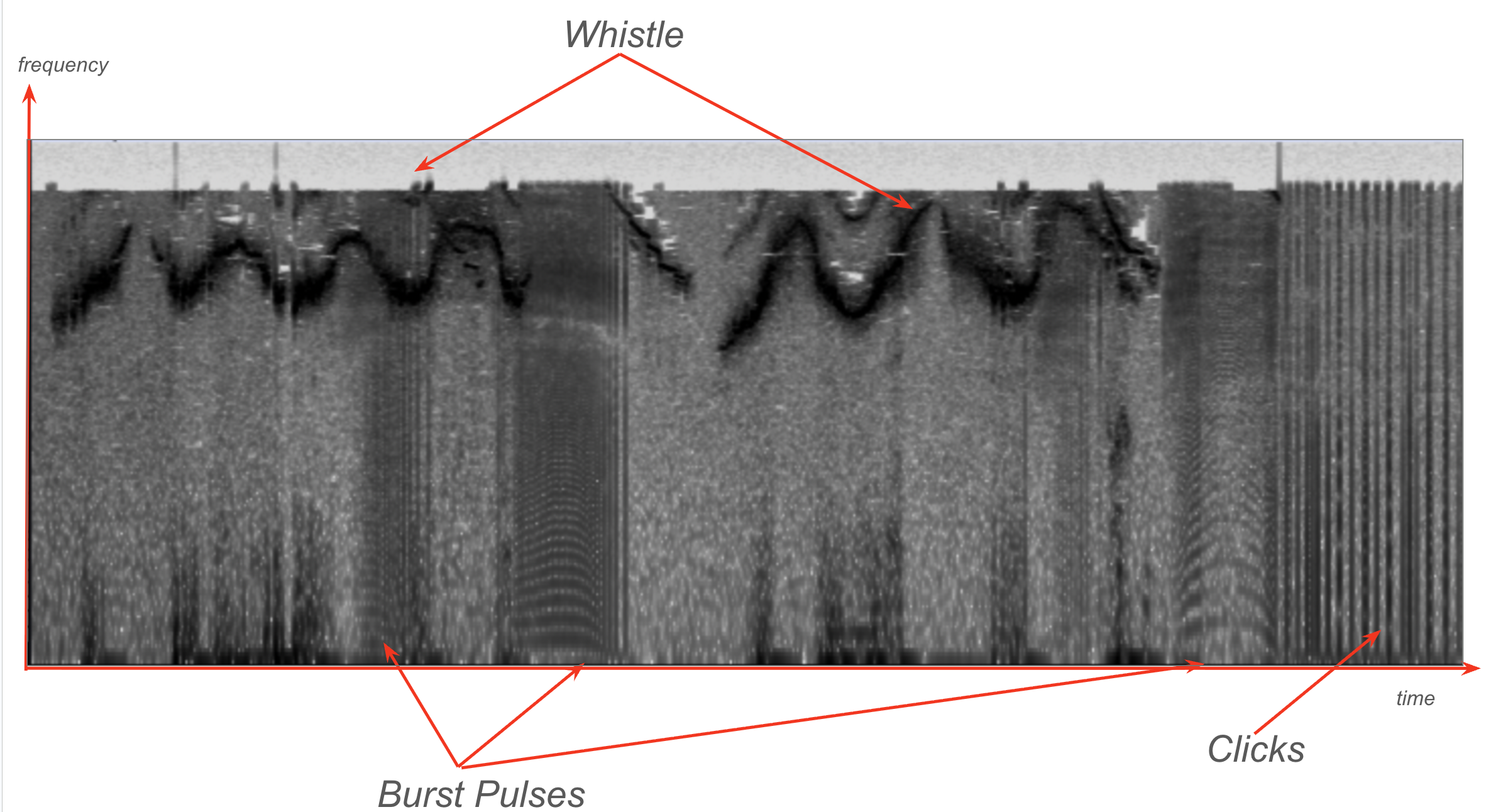}
\caption{A spectrogram with whistles, burst pulses and echo location clicks.}
\label{fig:spec}
\end{figure}

In order to visualize audible dolphin communication data, marine mammalogists look at the spectrogram of an audio recording. Mathematically, this spectrogram is represented as a sequence $S = \{ s_1, \dots, s_T \}$, where each $s_t$ is a spectral feature vector at time $t$. Using visual inspection and manual measurements, biologists have created various categorizations for these audible signals. One primary category identified by researchers is the dolphin whistle, which is conceptualized as a single oscillator changing frequency over time. In a spectrogram, a whistle appears as a single line where each point represents the oscillator's frequency at a specific moment. Another common category is the burst-pulse signal, consisting of a dense series of loud clicks. These appear in the spectrogram as multiple parallel lines, distinct from the singular wavy path of a whistle. Beyond these communicative signals, dolphins also utilize echolocation clicks. These clicks are characterized by sharp, vertical broadband pulses that span a wide range of frequencies almost instantaneously. Unlike the sustained tone of a whistle or the rhythmic grouping of a burst pulse, echolocation clicks appear as distinct, thin vertical lines that represent the dolphin's biological sonar used for navigation and hunting (see Figure \ref{fig:spec}).

\section{Alignment Algorithms}
In the following sections, we will introduce the concept of alignments, multiple sequence alignments (or profiles) and how to construct them. We will also explain how to implement the methods for dolphin communication. Since a deep introduction into the topic is beyond the scope of this paper, we recommend a classic text \cite{durbin} as an introduction. 

\begin{figure}[ht]
\centering
\includegraphics[width=0.9 \linewidth]{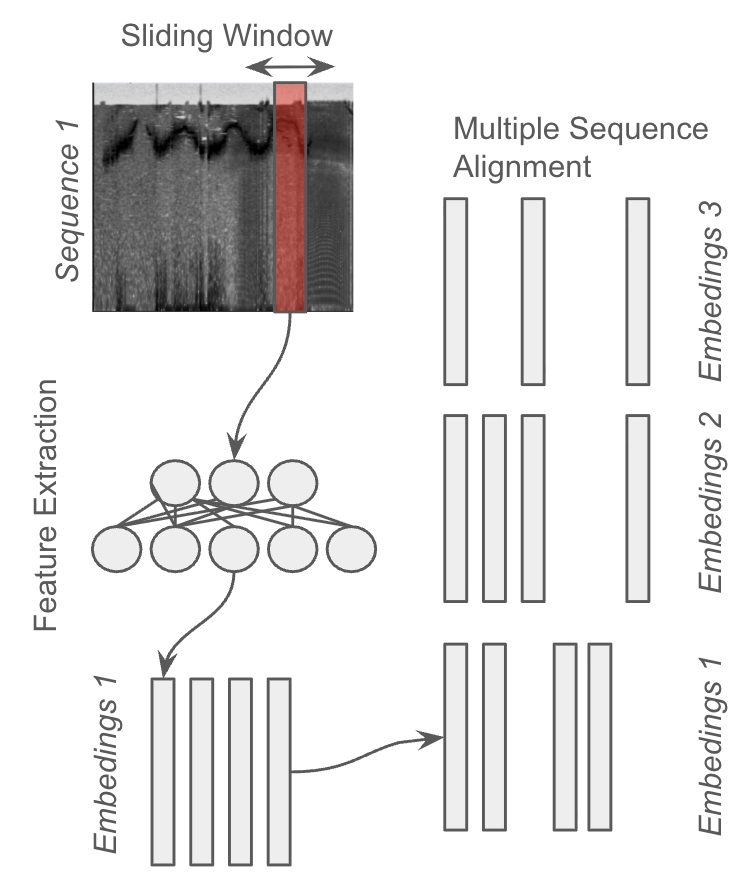}
\caption{Overview of the acoustic feature extraction and alignment pipeline. A sliding window moves across the spectrogram of an input recording (Sequence 1), extracting short, overlapping acoustic segments (red highlight). Each segment is passed through the feature extraction network (Section \ref{sec:featrues}) to produce a single embedding vector, and the embeddings for all windows in a recording are concatenated into an ordered embedding sequence (Embeddings 1). This process is repeated independently for every recording in the dataset, yielding one embedding sequence per input (Embeddings 1, 2, 3, ...). These embedding sequences are then passed to the multiple sequence alignment stage (Section \ref{sec:msa}), which progressively aligns them into a shared column structure, inserting gaps where necessary so that acoustically corresponding frames across recordings occupy the same column.}
\label{fig:pipeline}    
\end{figure}
Figure \ref{fig:pipeline} summarizes the full pipeline used to transform raw dolphin vocalization recordings into the multiple sequence alignments presented in this paper. Each input recording is first processed with a sliding window over its spectrogram, producing a series of short, overlapping acoustic segments that capture local spectral content at a fine temporal resolution. Every windowed segment is independently passed through our transformer-based feature extractor, which projects the high-dimensional spectrogram content into a compact embedding vector. Concatenating the embeddings for all windows in a recording, in temporal order, yields a single embedding sequence that serves as a continuous-valued analogue to a symbolic string in classical bioinformatics sequence alignment. This procedure is applied independently to every recording in the dataset, producing one embedding sequence per recording. The resulting set of embedding sequences is then handed to the progressive multiple sequence alignment algorithm, which uses the Gaussian kernel similarity measure in place of discrete substitution scores to iteratively align sequences and sequence profiles, ultimately producing the shared column structure used for downstream visualization and clustering.

To support reproducibility, our full implementation—including the
feature extraction, alignment, and clustering code described in
this section—is available at
\url{https://github.com/dkohlsdorf/dolphin_msa}.

\subsection{What is an Alignment}
Sequence Alignment is the process of re-arranging two or more sequences to reveal regions of similarity. In practice, this means plotting sequences on a shared grid so that identical patterns occupy the same vertical columns. 

\begin{table}[h]
\centering
\texttt{ % Encloses the table in typewriter font
\begin{tabular}{ll}
Column:     & 0123456789 \\
sequence 1: & *******\_\_\_ \\
sequence 2: & ********** \\
sequence 3: & *\_*\_\_\_****\\
sequence 4: & *\_*\_\_\_**** \\
sequence 5: & ********** \\
\end{tabular}
}
\caption{Multiple sequence alignment showing gaps (\_) used for local stretching.}
\end{table}
To achieve these identical vertical columns, alignment algorithms introduce gaps $(\_)$ into the sequences. These gaps act as "flexible spacers" that stretch or shrink a sequence locally. This technique ensures that "motifs" (functional patterns) align horizontally across the dataset, even if they occurred at different indices in the raw data. In this view, every position is a column where each sequence contributes either its original value or a gap.
If we align more than two sequences the process is called a multiple sequence alignment (MSA) or profile in the bio-informatics literature.

In the following sections, we will first explain how to align two sequences (pairwise alignment) and then switch to multiple alignments. Finally, we will adopt these methods to audio data.

\subsection{Constructing Pairwise Alignments}
There is a class of algorithms that use dynamic programming to align two sequences. For continuous sequences, the main algorithm is called dynamic time warping \cite{sakoe1978} which is frequently used in speech and signal processing as well as time series mining. The counterpart for discrete sequences such as text is called the Levenshtein distance. Both minimize the distance between the aligned sequences. Finally, the bio informatics algorithm with the same style of recursion is called Needleman-Wunsch \cite{NEEDLEMAN1970443}, which maximizes the alignment score. 

Formally, two sequences $Q = q_1 .... q_n$ and $C=c_1 .. c_m$ can be aligned by the following recursion: 

\begin{math}
D(i, j) = \text{opt} \begin{cases} 
D(i-1, j-1) + \gamma_{\text{diag}} \\
D(i-1, j) + \gamma_{\text{vert}} \\ 
D(i, j-1) + \gamma_{\text{horiz}} 
\end{cases}
\end{math}

\begin{figure}[ht]
\centering
\includegraphics[width=0.5\linewidth]{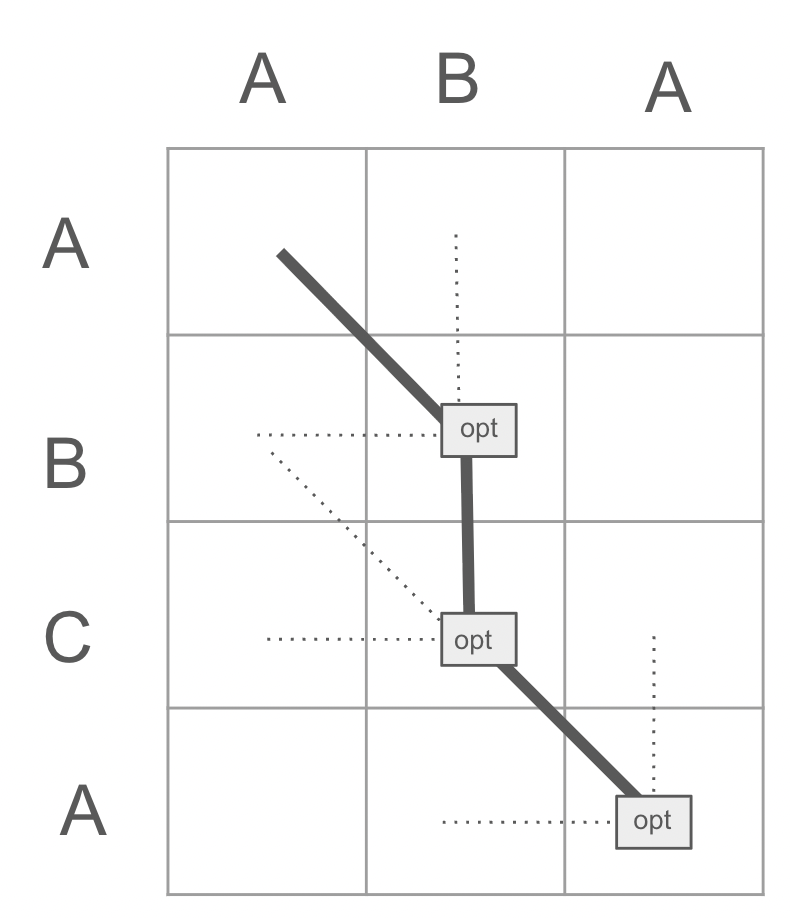}
\caption{A standard dynamic programming matrix for pairwise sequence alignment. The solid path highlights the optimal alignment between two individual sequences, where each step represents either a character match/mismatch (diagonal) or a single gap insertion (horizontal/vertical) based on a localized scoring reward system}
\label{fig:placeholder}
\end{figure}

\begin{itemize}
\item \textbf{Dynamic Time Warping (DTW):} The scoring function is the distance metric $d(q_i, c_j)$ and is typically the Euclidean distance. The alignment is constructed by finding the \textbf{minimum} cumulative distance along the alignment path, where diagonal, vertical, and horizontal steps are weighted equally by the local distance.

\item \textbf{Levenshtein Distance:} The scoring function is defined by unit edit costs. The alignment is constructed as the \textbf{minimum} number of operations required to transform one sequence into another, where a diagonal step costs $0$ for an exact match and $1$ for a substitution, while vertical and horizontal steps (insertions and deletions) always cost $1$.

\item \textbf{Needleman-Wunsch:} The scoring function utilizes a similarity matrix $s(q_i, c_j)$ and a constant gap penalty $g$. The alignment is constructed by seeking the \textbf{maximum} total score, where diagonal moves represent character matches or substitutions and vertical/horizontal moves represent penalized gaps.
\end{itemize}

The parameters for each algorithm are shown in Table \ref{tab:alignment-params}.

\begin{table}[ht]
\centering
\begin{tabular}{lcccc}
\toprule
\textbf{Algorithm} & \textbf{opt} & $\gamma_{\text{diag}}$ & $\gamma_{\text{vert}}$ & $\gamma_{\text{horiz}}$ \\ 
\midrule
DTW & $\min$ & $d(q_i, c_j)$ & $d(q_i, c_j)$ & $d(q_i, c_j)$ \\ \addlinespace
Levenshtein & $\min$ & $0 \text{ if } q_i = c_j \text{ else } 1$ & $1$ & $1$ \\ \addlinespace
Needlemann & $\max$ & $s(q_i, c_j)$ & $g$ & $g$ \\ 
\bottomrule
\end{tabular}
\caption{Unified Parameters for Sequence Alignment Algorithms}
\label{tab:alignment-params}
\end{table}

\subsection{Constructing Multiple Sequence Alignments and Progressive Alignment}
\label{sec:msa}
The construction of a Multiple Sequence Alignment (MSA) is a multi-stage process that leverages the mathematical rigor of pairwise alignment to build a complex, multi-sequence grid. The "guide tree" acts as the architectural blueprint for this construction, ensuring that the most biologically (or acoustically) similar sequences are merged first to minimize the introduction of alignment artifacts.
Before the tree is built, the algorithm performs all-against-all pairwise local alignments. For each pair of sequences, a similarity score is calculated using dynamic programming—such as the Needleman-Wunsch or Dynamic Time Warping algorithms. This process results in a distance matrix where each entry represents the cost or similarity between sequence $i$ and sequence $j$.

The distance matrix serves as the foundation for hierarchical clustering, typically through methods like UPGMA \cite{upgma} or Neighbor-Joining \cite{njtree}, which effectively "bakes" the clustering relationships directly into the alignment. In this structure, the leaves represent individual vocalization sequences, while the clustering logic dictates that sequences with the smallest distance—and thus the highest similarity—are joined first to form the initial branches of the guide tree. As the tree develops, each node represents a consensus "profile" of the sequences nested beneath it. Following this guide tree from the branches toward the root, the algorithm executes a series of progressive merges: first by aligning the most similar sequence-to-sequence pairs, then by performing sequence-to-profile alignments where additional sequences are added to existing consensus groups, and finally through profile-to-profile merges. In these final stages, two existing alignments are treated as single entities, and gaps are inserted as flexible spacers to preserve the alignment of shared motifs across the entire set.

\begin{figure}[ht]
\centering
\begin{forest}
for tree={
font=\ttfamily,
draw,
align=center,
child anchor=north,
parent anchor=south,
l sep=0.25cm,
s sep=.5cm
}
[\_\_bb\\a\_bb\\aaab\\aaaa
[a\_bb\\aaab\\aaaa
  [aaaa\\aaab
    [aaaa]
    [aaab]
  ]
  [abb]
]
[bb]
]
\end{forest}
\caption{A guide tree for progressive alignment. The process begins at the leaves with individual sequences and merges them into clusters (internal nodes) until a full Multiple Sequence Alignment is formed at the root.}
\label{fig:guide_tree}
\end{figure}
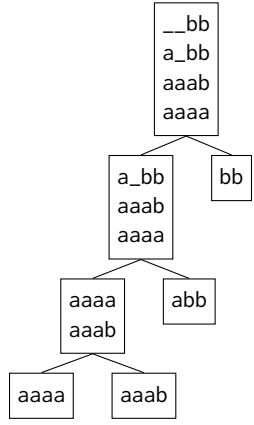

By the time the root of the tree is reached, every sequence has been plotted onto a shared grid where each vertical position is a column of feature vectors in time. This ensures that even if specific acoustic motifs occurred at different indices in the raw recordings, they are aligned horizontally for comparative analysis.

\begin{figure}[ht]
\centering
\includegraphics[width=0.8\linewidth]{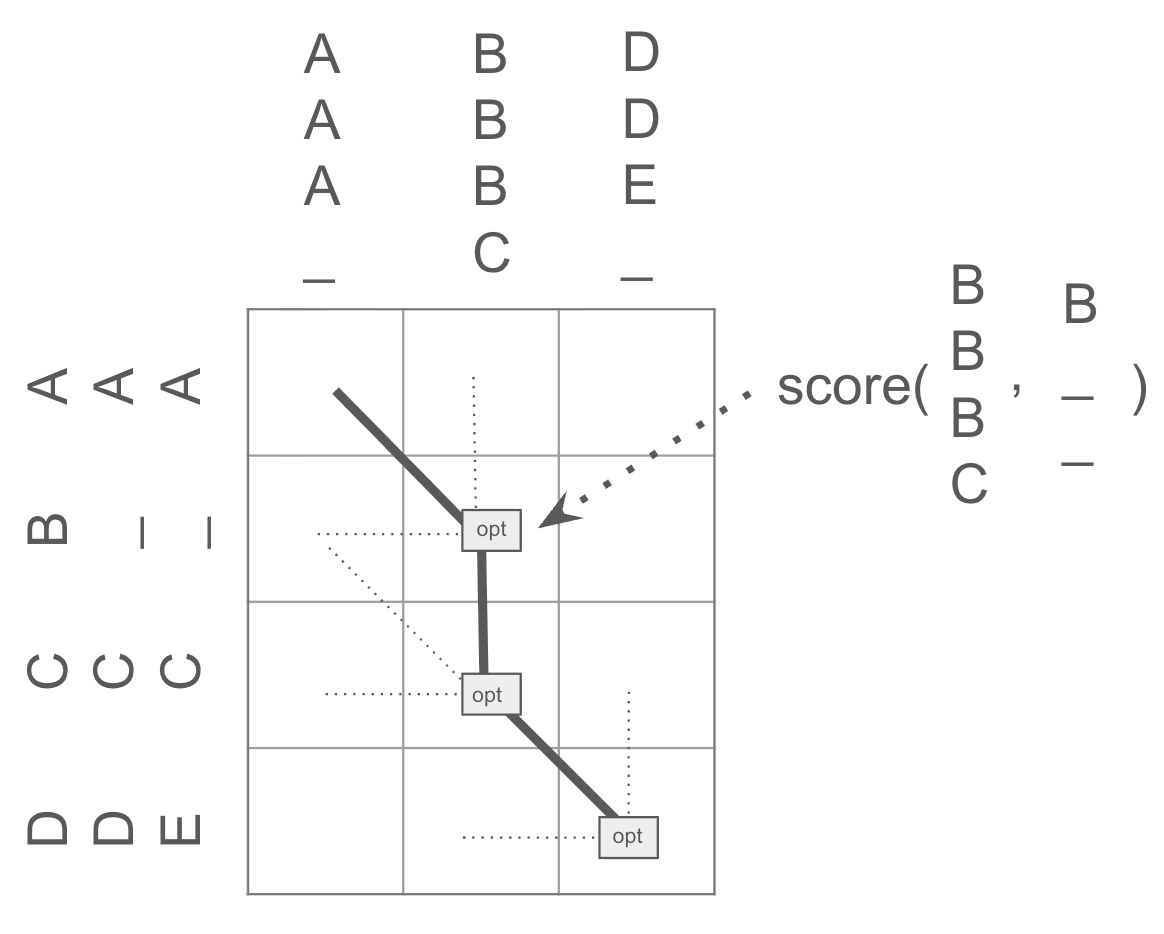}
\caption{A dynamic programming matrix for Profile-to-Profile alignment. The top profile consists of four sequences: $ABD$, $ABD$, $ABE$, and $\_C\_$.  The profile on the left consists of three sequences: $ABCD$, $A\_CD$, and $A\_CE$.  The thick black line represents the optimal trace-back path, where each "opt" node is calculated using a scoring function—such as Sum-of-Pairs—to evaluate the cost of aligning multi-sequence columns, effectively merging two sub-alignments into a unified progressive alignment.  A $\_$ will be inserted between the second and third letters in the four top sequences to make all seven sequences four letters long ($AB\_D$, $AB\_D$, $AB\_E$, $\_C\_\_$, $ABCD$, $A\_CD$, and $A\_CE$)}
\label{fig:msa_paths_simple}
\end{figure}

To implement the progressive alignment, we can utilize the same underlying class of algorithms used for pairwise alignment—such as Needleman-Wunsch or Dynamic Time Warping (DTW)—but adapt them for Sequence-to-Profile and Profile-to-Profile alignments (see Figure \ref{fig:msa_paths_simple}). Instead of aligning raw sequences of individual characters or acoustic frames, the algorithm aligns columns of data. In this framework, a single sequence is treated as an alignment with only one row, while a profile represents an existing alignment containing multiple rows of sequences and gaps. When two profiles are merged, the scoring function evaluates the similarity between columns of feature vectors rather than individual points. 
In discrete sequence alignment, column scoring typically relies on two primary methodologies: Identity Consensus and Sum-of-Pairs. Identity Consensus focuses on matching characters based on the most frequent symbol within a specific column, incorporating specific penalties whenever gaps are introduced to ensure the structural integrity of the alignment. Alternatively, the Sum-of-Pairs approach offers a more comprehensive metric by calculating the substitution cost between every possible pair of characters within the two columns, providing a detailed measure of cumulative similarity across the entire aligned set.

Formally, given two alignment columns $C_x$ and $C_y$, each containing symbols from an alphabet $\Sigma$ (or the gap symbol $-$), the Sum-of-Pairs score is computed as:
\begin{equation}
S_{\text{SOP}}(C_x, C_y) = \sum_{a \in C_x} \sum_{b \in C_y} s(a, b)
\end{equation}
where $s(a,b) = +\varepsilon$ if $a = b$ and $a \neq -$ (matching non-gap symbols), and $s(a,b) = -\varepsilon$ otherwise. This sum can be efficiently computed by counting symbol occurrences. Let $n_x^i$ and $n_y^i$ denote the count of symbol $i$ in columns $C_x$ and $C_y$ respectively. The score simplifies to:
\begin{equation}
S_{\text{SOP}}(C_x, C_y) = \frac{2\varepsilon \cdot \sum_{i \in \Sigma} n_x^i \cdot n_y^i}{|C_x| \cdot |C_y|} - \varepsilon
\end{equation}
This formulation reduces the computational complexity from $O(|C_x| \cdot |C_y|)$ to $O(|\Sigma|)$ by leveraging the distributive property of the pairwise summation over symbol counts.

In the next section, we will describe how to adopt this to continuous data such as audio. While the progressive alignment logic remains the same, the data structure shifts from characters to columns of feature vectors in time. By treating each time-step as a multidimensional vector, we can apply the same hierarchical clustering and guide-tree logic to find consensus patterns in complex dolphin vocalizations.

\subsubsection{Early Stopping and Multiple Alignment Clusters}
\label{sec:early-stopping}
The progressive alignment procedure described above continues merging until all sequences have been incorporated into a single multiple sequence alignment rooted at the top of the guide tree. However, when analyzing large and heterogeneous datasets, such as dolphin vocalizations recorded across different behavioral contexts, forcing all sequences into a single alignment may obscure meaningful distinctions between structurally dissimilar groups. Sequences representing aggressive burst pulses, for instance, share little temporal structure with feeding whistles, and aligning them together would introduce excessive gaps and dilute the conserved patterns within each category.

To address this limitation, we introduce an early stopping criterion based on the distance matrix used to construct the guide tree. Rather than merging all the way to the root, we halt the progressive alignment when the distance between candidate clusters exceeds a specified threshold. This halting produces multiple independent alignments, each containing sequences that share sufficient structural similarity to warrant direct comparison.

We support two complementary stopping criteria. The first approach uses a distance percentile threshold: given the distribution of all pairwise distances in the dataset, we compute a cutoff at a specified percentile (e.g., the 50th percentile) and refuse to merge clusters whose distance exceeds this value. This adaptive threshold automatically adjusts to the natural variability within each dataset.

The second approach specifies a target number of clusters directly. The hierarchical clustering proceeds using a chosen linkage method, such as UPGMA (Unweighted Pair Group Method with Arithmetic Mean), WPGMA (Weighted Pair Group Method), complete linkage, or neighbor-joining, and terminates when the desired number of clusters is reached. Each resulting cluster is then aligned independently using the progressive alignment algorithm, producing a set of multiple sequence alignments rather than a single monolithic structure.

This hierarchical clustering approach offers several advantages for exploratory analysis. Researchers can examine alignments at multiple granularities by varying the distance threshold or target cluster count, revealing structure at different levels of similarity. The resulting cluster assignments also provide natural groupings for downstream analysis, such as identifying behavioral correlates or comparing vocalization patterns across social contexts.

\subsection{Dolphin Vocalization Alignment}
While the discrete Sum-of-Pairs scoring function works well for symbolic sequences such as DNA or protein chains, dolphin vocalizations present a fundamentally different challenge. When marine mammalogists examine a spectrogram, they observe a continuous stream of acoustic energy distributed across frequency bands over time. Each time step in a spectrogram is not a single character from a finite alphabet, but rather a high-dimensional vector capturing the spectral energy distribution at that moment. Consequently, the notion of an ``exact match'' between two acoustic frames becomes meaningless; two dolphin whistles may be highly similar without ever being numerically identical.

To address this challenge, we replace the discrete substitution score with a continuous similarity measure based on the Gaussian kernel \cite{vert2004primer}. The intuition behind this approach is straightforward: two acoustic frames that are close together in the feature space should receive a high alignment score, while frames that are far apart should be penalized. The Gaussian kernel provides a natural way to quantify this notion of proximity, producing similarity values that decay smoothly as the distance between vectors increases. When two spectral frames are nearly identical, their similarity approaches one; when they are vastly different, their similarity approaches zero.

\begin{figure}[ht]
\centering
\includegraphics[width=1.0\linewidth]{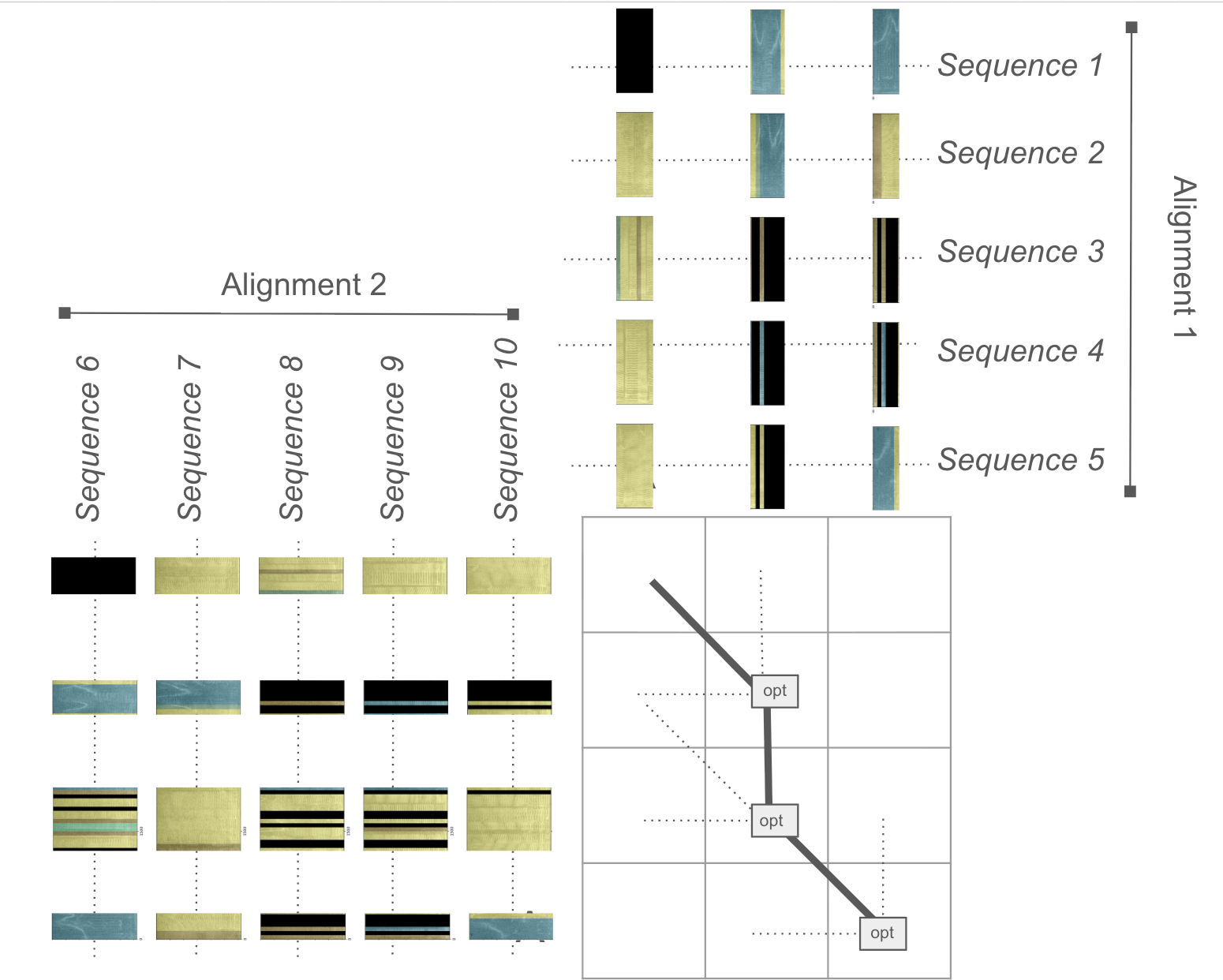}
\caption{A multiple sequence alignment visualization demonstrating the process of aligning two pre-aligned spectrogram profiles (Alignment 1, vertical; Alignment 2, horizontal). Individual sequence thumbnails (chromagram or Mel-spectrogram segments) show acoustic data with inserted black 'gaps' to maintain internal alignment. A cost matrix with an optimal dynamic path ('opt' nodes) determines the final combined alignment structure}
\label{fig:msa_paths}
\end{figure}

Consider the alignment of two spectrogram profiles as illustrated in Figure~\ref{fig:msa_paths}. Each profile consists of multiple sequences that have already been aligned internally, with gaps (shown as black regions) inserted to maintain temporal correspondence. When merging these two profiles, the algorithm must evaluate how well each column from the first profile matches each column from the second. Unlike the discrete case where we simply count matching symbols, we must now assess the aggregate similarity between two sets of continuous feature vectors.

Formally, let $C_x = \{\mathbf{v}_1, \mathbf{v}_2, \ldots, \mathbf{v}_k\}$ and $C_y = \{\mathbf{u}_1, \mathbf{u}_2, \ldots, \mathbf{u}_l\}$ denote two alignment columns, where each $\mathbf{v}_i, \mathbf{u}_j \in \mathbb{R}^d$ represents a $d$-dimensional spectral feature vector (or is marked as a gap). The pairwise similarity between any two non-gap vectors is computed using the Gaussian kernel:
\begin{equation}
\text{sim}(\mathbf{v}_i, \mathbf{u}_j) = \exp\left(-\frac{\|\mathbf{v}_i - \mathbf{u}_j\|^2}{2\sigma^2}\right)
\end{equation}
where $\sigma$ is a bandwidth parameter that controls the sensitivity of the similarity measure. A small $\sigma$ produces sharp distinctions between similar and dissimilar frames, while a larger $\sigma$ yields a more gradual transition. In practice, we set $\sigma$ based on the average pairwise distance observed in the dataset, ensuring that the similarity measure is calibrated to the natural variability of dolphin vocalizations.

To obtain the column score, we compute the average similarity across all valid (non-gap) pairs of vectors from the two columns. Let $C_x'$ and $C_y'$ denote the subsets of $C_x$ and $C_y$ after removing gaps. The average pairwise similarity is then:
\begin{equation}
\bar{s}(C_x, C_y) = \frac{1}{|C_x'| \cdot |C_y'|} \sum_{\mathbf{v} \in C_x'} \sum_{\mathbf{u} \in C_y'} \text{sim}(\mathbf{v}, \mathbf{u})
\end{equation}
This average similarity lies in the interval $[0, 1]$, where values close to one indicate that the acoustic frames in both columns are tightly clustered in feature space, and values close to zero indicate substantial divergence. To convert this similarity into an alignment score compatible with the dynamic programming framework, we apply a linear transformation:
\begin{equation}
S(C_x, C_y) = \varepsilon \cdot (2\bar{s}(C_x, C_y) - 1)
\end{equation}
where $\varepsilon$ is a scaling parameter that determines the magnitude of match and mismatch scores. This transformation maps the similarity range $[0, 1]$ onto the score range $[-\varepsilon, +\varepsilon]$. When columns contain highly similar acoustic content, the score approaches $+\varepsilon$, rewarding the alignment. When columns are dissimilar, the score approaches $-\varepsilon$, penalizing the match and encouraging the algorithm to insert gaps instead.

The dynamic programming procedure for aligning two profiles proceeds exactly as in the discrete case, with the Gaussian-based column score replacing the Sum-of-Pairs function. Figure~\ref{fig:msa_paths} illustrates this process for two spectrogram profiles. The horizontal axis represents the columns of one profile, while the vertical axis represents the columns of the other. Each cell in the cost matrix accumulates the optimal alignment score up to that point, and the traceback path (indicated by the ``opt'' nodes) reveals the final correspondence between columns. Where the path moves diagonally, columns from both profiles are merged; where it moves horizontally or vertically, gaps are inserted into one profile to accommodate insertions or deletions in the acoustic stream.

This continuous scoring framework preserves the essential structure of the ClustalW progressive alignment algorithm while adapting it to the realities of acoustic data. The resulting multiple sequence alignments reveal temporal correspondences across dolphin vocalizations, highlighting repeated motifs and structural patterns that would be difficult to detect through simple spectrogram inspection alone.

\subsection{Feature extraction using transformers}
\label{sec:featrues}

While the Gaussian kernel scoring function described in Section 5.4 provides a principled approach to comparing acoustic frames, directly applying it to raw spectrogram representations presents significant computational and statistical challenges. A typical spectrogram frame consists of 80 mel-frequency bins, and when considering temporal context, the effective dimensionality quickly escalates to thousands of parameters. In such high-dimensional spaces, the curse of dimensionality renders distance metrics increasingly unreliable—all points become approximately equidistant, and the Gaussian kernel's ability to distinguish between similar and dissimilar frames degrades substantially.

To address this fundamental limitation, we employ a transformer-based feature extractor that projects raw spectrograms into a compact, semantically meaningful embedding space. Specifically, we adapt OpenAI's Whisper model \cite{radford2022whisper}, a large-scale audio encoder pre-trained on diverse human speech data, to the domain of dolphin vocalizations. Although Whisper was designed for automatic speech recognition, its encoder has learned rich acoustic representations that transfer effectively to non-human vocalization analysis.

We fine-tune Whisper's encoder using Low-Rank Adaptation (LoRA) \cite{hu2021lora}, which introduces trainable low-rank decomposition matrices into the attention and feed-forward layers while keeping the original pre-trained weights frozen. This approach enables efficient domain adaptation with minimal additional parameters—we apply LoRA with rank 16 to the query, key, value, and output projection matrices, as well as the feed-forward layers. The fine-tuning procedure optimizes a composite objective combining two complementary loss functions.

The first component is a triplet loss \cite{jensen_tripletloss} that enforces metric learning constraints on the embedding space. Given an anchor acoustic segment, a positive example from the same vocalization type, and a negative example from a different type, the triplet loss encourages the model to produce embeddings where acoustically similar signals cluster together while dissimilar signals are pushed apart. The second component is a classification loss that trains the model to distinguish between vocalization categories (whistles, burst pulses, echolocation clicks, and noise), providing an additional supervisory signal that shapes the embedding geometry according to biologically meaningful distinctions.

The fine-tuned encoder feeds into a projection head that produces 128-dimensional embeddings for each acoustic window. This dimensionality reduction—from thousands of spectrogram coefficients to a compact 128-dimensional vector—dramatically improves the reliability of similarity computations while preserving the acoustic information essential for alignment. The resulting embeddings exhibit the property that vocalizations with similar acoustic structure map to nearby points in the embedding space, enabling the Gaussian kernel scoring function to operate effectively.

Audio is processed in overlapping windows of approximately 320 milliseconds, with each window bandpass-filtered to the dolphin vocalization range (5–10 kHz) before feature extraction. The model additionally produces classification probabilities that enable noise filtering: windows classified as noise with high confidence are excluded from subsequent alignment, ensuring that the multiple sequence alignment focuses on communicative signals rather than ambient underwater sounds. 

\subsection{Column Analysis and Clustering}
\label{sec:clustering}
Once a multiple sequence alignment has been constructed, the resulting grid of aligned acoustic frames provides a powerful scaffold for comparative analysis. However, interpreting these alignments visually remains challenging: each column contains multiple embedding vectors from different recordings, and understanding the acoustic content at each position requires examining high-dimensional feature representations. To facilitate interpretation, we assign discrete cluster labels to each acoustic frame, enabling color-coded visualizations where similar sounds share the same hue across the alignment (see Figure 1).

Standard clustering algorithms such as k-means or Gaussian mixture models operate on the embedding space independently of the alignment structure. While these methods can identify acoustically similar frames, they ignore a crucial source of information: frames that have been aligned to the same MSA column are, by construction, candidates for representing the same underlying acoustic motif. This observation motivates an alignment-informed clustering approach that leverages the column structure as a prior to stabilize cluster assignments.

We adopt spectral clustering \cite{ng-spectral-clustering-analysis-2002} as our base algorithm, which operates by constructing an affinity matrix that captures pairwise similarities between all embedding vectors, then performing eigendecomposition to obtain a low-dimensional spectral embedding suitable for clustering. The key innovation lies in how we construct this affinity matrix. Given $N$ embedding vectors $\mathbf{e}_1, \ldots, \mathbf{e}_N$ collected across all aligned sequences, we first compute pairwise Euclidean distances and transform them into similarities using a radial basis function (RBF) kernel:

$$W_{ij} = \exp\left(-\frac{\mathbf{e}_i - \mathbf{e}_j^2}{\tau}\right)$$

where $\tau$ is a bandwidth parameter controlling the kernel's sensitivity. This process produces an affinity matrix where acoustically similar embeddings receive high similarity scores regardless of their position in the alignment.

To incorporate the MSA column structure, we augment this affinity matrix with a column prior. Let $\mathcal{C}(i)$ denote the MSA column to which embedding $i$ belongs. For each pair of embeddings $(i, j)$ that share the same column—that is, $\mathcal{C}(i) = \mathcal{C}(j)$—we add a constant boost $\lambda_{\text{col}}$ to their affinity:

$$W'{ij} = W{ij} + \lambda_{\text{col}} \cdot \mathbf{1}[\mathcal{C}(i) = \mathcal{C}(j)]$$

where $\mathbf{1}[\cdot]$ is the indicator function. This modification encourages embeddings aligned to the same column to cluster together, effectively using the alignment as soft supervision for the clustering process. The intuition is that the progressive alignment algorithm has already identified these frames as temporally corresponding across different recordings; the column prior propagates this structural information into the clustering objective.

After constructing the modified affinity matrix, we compute the normalized graph Laplacian and extract its top $k$ eigenvectors to obtain a spectral embedding. We then fit a Bayesian Gaussian mixture model \cite{bayesiangauss} to this low-dimensional representation, which automatically determines the effective number of clusters by pruning components with negligible mixing weights. This approach avoids the need to specify the exact number of clusters a priori while still providing probabilistic cluster assignments.

To further improve temporal coherence within individual recordings, we apply hidden Markov model (HMM) \cite{Rabiner1989} smoothing to the cluster assignments. The HMM enforces the prior that cluster labels should persist over consecutive frames rather than switching rapidly, which would produce visually fragmented and difficult-to-interpret alignments. The transition matrix is parameterized by an expected segment length $m_{\text{cluster}}$, with self-transition probability set to $(m_{\text{cluster}} - 1)/m_{\text{cluster}}$ to encourage smooth label sequences.

The resulting cluster assignments can be visualized as colored overlays on the aligned spectrograms, where each cluster receives a distinct color from a categorical palette. This visualization immediately reveals structural patterns: repeated motifs appear as consistent color bands across multiple sequences in the same columns, while acoustically distinct segments are differentiated by contrasting colors. The column-informed clustering ensures that these color assignments respect the alignment structure, producing coherent visualizations where the horizontal correspondence established by the MSA is reinforced by consistent vertical coloring.

\section{Experiments}
We evaluate our multiple sequence alignment framework through two complementary experiments designed to assess both technical performance and biological relevance. The first experiment examines the impact of transformer model capacity on alignment quality, comparing embeddings produced by Whisper models of different sizes to determine whether increased capacity yields more informative representations for acoustic alignment. The second experiment applies the framework to a developmental comparison between adult and juvenile dolphin vocalizations, testing the hypothesis that mature dolphins produce more stereotyped calls than juveniles still acquiring vocal control.

Both experiments operate on curated datasets of dolphin vocalizations extracted from video-annotated behavioral episodes collected by the Wild Dolphin Project. We apply identical preprocessing and alignment procedures to each dataset, enabling direct comparison of the resulting metrics. To quantify alignment quality, we introduce a comprehensive evaluation framework spanning gap structure, column variance, and clustering consistency---metrics that collectively characterize how well the algorithm identifies temporal correspondences across recordings.

The following subsections describe the datasets, define our evaluation metrics, and present results for each experiment. We report both quantitative metrics and qualitative observations from the resulting MSA visualizations, demonstrating that the framework produces interpretable alignments that reveal structural patterns consistent with domain knowledge about dolphin vocal behavior.
\subsection{Datasets}
We evaluate our alignment framework on two curated datasets of dolphin vocalizations collected by the Wild Dolphin Project, a long-term research initiative studying free-ranging Atlantic spotted dolphins (\textit{Stenella frontalis}) in the Bahamas. The original recordings span more than 30 summers of fieldwork, during which researchers captured synchronized underwater video and audio of dolphin social interactions across diverse behavioral contexts.

Marine mammalogists manually reviewed video recordings to identify and annotate behavioral episodes, including aggressive encounters between adults and play behavior among juveniles. For each annotated episode, the corresponding audio segment was extracted from the synchronized recording, producing isolated vocalization snippets suitable for acoustic analysis. This manual curation process ensures that each audio file captures vocalizations produced during a specific, verified behavioral context, providing ground-truth labels for our comparative analysis.

The adult aggression dataset comprises 58 audio recordings extracted from aggressive behavioral episodes involving adult dolphins. Aggressive encounters are characterized by synchronized burst pulses---dense series of broadband clicks that appear as multiple parallel lines in the spectrogram---often accompanied by physical displays such as head-to-head posturing or chasing. The recordings originate primarily from two extended observation sessions in 2021, with file durations ranging from approximately 0.2 to 2.5 seconds per vocalization snippet.

The juvenile play aggression dataset contains 96 audio recordings of play aggression vocalizations produced by juvenile dolphins. Play behavior in young dolphins often mimics adult aggression patterns but occurs in non-threatening social contexts, typically involving age-matched peers. The recordings span multiple field seasons from 2020 to 2023 and capture a broader range of individual variation, reflecting both the developmental diversity within the juvenile cohort and the extended observation period. File durations range from approximately 0.05 to 3.2 seconds.

\begin{table}[h]
\centering
\caption{Summary of evaluation datasets. Both datasets contain manually extracted vocalization snippets from video-annotated behavioral episodes.}
\label{tab:datasets}
\begin{tabular}{lrr}
\toprule
\textbf{Characteristic} & \textbf{Adult Aggression} & \textbf{Juvenile Play} \\
\midrule
Total recordings & 58 & 96 \\
Recording sessions & 2 & 6 \\
Collection period & 2021 & 2020--2023 \\
Duration range (s) & 0.2--2.5 & 0.05--3.2 \\
Behavioral context & Adult aggression & Juv. play aggression \\
\bottomrule
\end{tabular}
\end{table}

We deliberately selected smaller, well-curated datasets that are tractable for visual inspection and manual validation of alignment results.

The contrast between these datasets enables a developmental comparison: adult aggression calls represent mature, presumably stereotyped vocalizations produced by individuals with full motor control over their vocal apparatus, while juvenile play calls capture the more variable acoustic output of young dolphins still acquiring vocal proficiency. By applying identical alignment procedures to both datasets, we can quantify differences in vocalization consistency that may reflect underlying developmental trajectories in cetacean vocal learning.

\subsection{Alignment Quality Metrics}
Evaluating the quality of a multiple sequence alignment requires examining several complementary aspects of the resulting structure. We organize our evaluation metrics into four categories: gap structure metrics that characterize how gaps are distributed throughout the alignment, per-sequence statistics that enable comparison across datasets, column quality metrics that assess the homogeneity of aligned positions, and clustering quality metrics that evaluate whether the column-informed clustering produces consistent assignments. In the following subsections, we define each metric and describe how it relates to alignment quality.

We formalize each metric introduced in this section. Let an alignment consist of $N$ sequences and $L$ columns, with entry $x_{ij}$ denoting the content of sequence $i$ at column $j$, where $x_{ij} = \mathrm{gap}$ if position $(i,j)$ is a gap and $x_{ij} = e_{ij} \in \mathbb{R}^d$ otherwise (the embedding vector of that acoustic frame).

\subsubsection{Gap Structure Metrics}
The distribution of gaps within an alignment reveals how much temporal stretching was required to bring sequences into correspondence. We first distinguish between \emph{gap openings} and \emph{gap extensions}: a gap opening occurs when a new gap region begins, while a gap extension continues an existing gap. The ratio of gap openings to gap extensions indicates the fragmentation of gaps throughout the alignment, a lower ratio suggests that gaps occur in longer, more contiguous blocks rather than being scattered throughout the sequences. Well-formed alignments typically exhibit ratios between 0.1 and 1.0.

The total number of gaps across all sequences provides a raw measure of alignment density, with fewer gaps indicating that sequences share more content in common. We normalize this quantity as the \emph{gap percentage}, which expresses gaps as a fraction of total alignment positions. Alignments with gap percentages between 10\% and 40\% generally represent a reasonable balance between accommodating sequence variation and maintaining meaningful correspondence. More formally, the total number of gaps in the alignment is
\begin{equation}
  \text{total\_gaps} = \sum_{i=1}^{N} \sum_{j=1}^{L} \mathbb{I}\!\left[x_{ij} = \mathrm{gap}\right],
\end{equation}
normalized as a percentage of all alignment positions:
\begin{equation}
  \text{gap\_percentage} = \frac{\text{total\_gaps}}{N \cdot L} \times 100.
\end{equation}
We distinguish \emph{gap openings} from \emph{gap extensions}. A gap
opening occurs at position $(i,j)$ if $x_{ij} = \mathrm{gap}$ and
either $j=1$ or $x_{i,j-1} \neq \mathrm{gap}$:
\begin{equation}
  \text{gap\_opens} = \sum_{i=1}^{N} \sum_{j=1}^{L}
    \mathbb{I}\!\left[x_{ij} = \mathrm{gap} \;\wedge\; \left(j=1 \;\vee\; x_{i,j-1} \neq \mathrm{gap}\right)\right].
\end{equation}
A gap extension occurs when a gap continues an already-open gap:
\begin{equation}
  \text{gap\_extends} = \sum_{i=1}^{N} \sum_{j=2}^{L}
    \mathbb{I}\!\left[x_{ij} = \mathrm{gap} \;\wedge\; x_{i,j-1} = \mathrm{gap}\right].
\end{equation}
The gap ratio, indicating gap fragmentation, is then
\begin{equation}
  \text{gap\_ratio} = \frac{\text{gap\_opens}}{\text{gap\_extends}}.
\end{equation}

\subsubsection{Per-Sequence Gap Statistics}
To enable meaningful comparison across datasets of different sizes, we compute gap statistics at the sequence level and then aggregate across the alignment. The mean and median gap percentages across all sequences indicate typical alignment quality, with the median providing robustness to outlier sequences that may contain unusually long silent regions. The standard deviation of gap percentages reveals whether alignment quality is uniform across sequences or whether certain recordings required substantially more gap insertion than others.

We additionally track the maximum gap percentage among all sequences to identify potential outliers, as well as the mean and maximum consecutive gap lengths. Long consecutive gap runs may indicate regions where a sequence has no corresponding content in the alignment, which can arise from recordings that capture only partial vocalizations or contain extended periods of silence.

Formally, for each sequence $i$, the per-sequence gap percentage is
\begin{equation}
  g_i = \frac{1}{L} \sum_{j=1}^{L} \mathbb{I}\!\left[x_{ij} = \mathrm{gap}\right] \times 100.
\end{equation}
We report the mean, median, standard deviation, and maximum of
$\{g_i\}_{i=1}^{N}$ across all sequences:
\begin{align}
  \text{mean\_gap\_pct} &= \frac{1}{N}\sum_{i=1}^{N} g_i,
  &\text{median\_gap\_pct} &= \operatorname{median}_i \left(g_i\right), \\
  \text{std\_gap\_pct} &= \sqrt{\frac{1}{N}\sum_{i=1}^{N} \left(g_i - \overline{g}\right)^2},
  &\text{max\_gap\_pct} &= \max_i g_i.
\end{align}

Let $r_i$ denote the length of the longest run of consecutive gap
positions within sequence $i$:
\begin{equation}
  r_i = \max_{1 \le j \le L} \left\{ k : x_{i,j} = x_{i,j+1} = \dots = x_{i,j+k-1} = \mathrm{gap} \right\}.
\end{equation}
We report
\begin{equation}
  \text{mean\_max\_consecutive} = \frac{1}{N}\sum_{i=1}^{N} r_i,
  \qquad
  \text{max\_consecutive} = \max_i r_i.
\end{equation}

\subsubsection{Column Quality Metrics}

Beyond gap structure, we assess whether the acoustic content within each alignment column is internally consistent. For each column, we compute the variance of the embedding vectors contributed by different sequences, excluding gap positions. The mean and median column variance across the entire alignment indicate how tightly clustered the aligned frames are in embedding space, lower variance suggests that the alignment has successfully grouped acoustically similar content.

We classify columns based on their variance into \emph{clean columns} (variance below 0.1), which represent well-aligned positions where all contributing sequences exhibit similar acoustic characteristics, and \emph{noisy columns} (variance above 1.0), which may indicate alignment errors or positions where sequences diverge substantially. The percentage of clean columns serves as an overall quality indicator, with higher values suggesting more successful alignment.

For column $j$, let $E_j = \{ e_{ij} : x_{ij} \neq \mathrm{gap} \}$
denote the set of non-gap embedding vectors contributed to that
column, with mean $\bar{e}_j = \frac{1}{|E_j|}\sum_{e \in E_j} e$.
The column variance is
\begin{equation}
  \operatorname{var}(j) = \frac{1}{|E_j|} \sum_{e \in E_j} \left\| e - \bar{e}_j \right\|^2.
\end{equation}
We report the mean and median of $\operatorname{var}(j)$ over all
columns $j = 1, \dots, L$. A column is classified as \emph{clean} if
$\operatorname{var}(j) < 0.1$ and \emph{noisy} if
$\operatorname{var}(j) > 1.0$:
\begin{equation}
  \text{percentage\_clean} = \frac{100}{L}\sum_{j=1}^{L} \mathbb{I}\!\left[\operatorname{var}(j) < 0.1\right]
\end{equation}

\begin{equation}
  \text{percentage\_noise} = \frac{100}{L}\sum_{j=1}^{L} \mathbb{I}\!\left[\operatorname{var}(j) > 1.0\right].
\end{equation}

\subsubsection{Alignment Statistics}

When applying hierarchical clustering prior to alignment, some recordings may remain as singletons that cannot be meaningfully aligned with other sequences. We report the number of unaligned files and the corresponding \emph{alignment rate}, the percentage of input recordings that were successfully incorporated into multi-sequence alignments. Higher alignment rates indicate that the dataset contains sufficient structural similarity for the algorithm to identify correspondences.

Let $N_{\text{total}}$ denote the number of input recordings prior to
clustering (Section~\ref{sec:early-stopping}), and let
$\text{unaligned\_files}$ denote the number of recordings that
remained singletons and were not incorporated into any multi-sequence
alignment. The alignment rate is
\begin{equation}
  \text{alignment\_rate} = \frac{N_{\text{total}} - \text{unaligned\_files}}{N_{\text{total}}} \times 100.
\end{equation}

\subsubsection{Clustering Quality Metrics}

Finally, we evaluate whether the column-informed spectral clustering produces assignments that respect the alignment structure. For each column, we compute the \emph{purity} of cluster labels among the non-gap entries, defined as the fraction of entries belonging to the most frequent cluster in that column. The mean and median purity across all columns indicate clustering consistency, with values approaching 1.0 suggesting that aligned positions receive the same cluster label across different sequences.

We complement purity with entropy-based measures that quantify the diversity of cluster labels within each column. Lower entropy indicates that columns contain predominantly a single cluster, while higher entropy suggests mixed assignments. The percentage of \emph{pure columns}, those where all non-gap entries share exactly the same cluster label, rovides a stringent measure of alignment-clustering agreement. Together, these metrics validate that the column prior successfully encourages consistent labeling of temporally corresponding acoustic events.

For column $j$, let $c_{ij}$ denote the cluster label assigned to
the non-gap entry $x_{ij}$, and let $p_{jk}$ denote the fraction of
non-gap entries in column $j$ assigned to cluster $k$:
\begin{equation}
  p_{jk} = \frac{1}{|E_j|} \sum_{i : x_{ij} \neq \mathrm{gap}} \mathbb{I}\!\left[c_{ij} = k\right].
\end{equation}
Column purity and entropy are then
\begin{equation}
  \text{purity}(j) = \max_{k} p_{jk},
  \qquad
  \text{entropy}(j) = -\sum_{k} p_{jk} \log p_{jk}.
\end{equation}
We report the mean and median of $\text{purity}(j)$ and
$\text{entropy}(j)$ over all columns, as well as the standard
deviation of purity, $\text{cluster\_std\_purity}$. A column is
\emph{pure} if $\text{purity}(j) = 1$, i.e., all non-gap entries in
that column share the same cluster label:
\begin{equation}
  \text{cluster\_pct\_pure\_cols} = \frac{100}{L}\sum_{j=1}^{L} \mathbb{I}\!\left[\text{purity}(j) = 1\right].
\end{equation}

\section{Experiments and Evaluation}
\label{sec:evaluation}

We evaluate our multiple sequence alignment (MSA) framework by analyzing how different acoustic feature representations and model configurations perform on adult aggression and juvenile play datasets. To quantify alignment performance, our evaluation framework measures gap structure, per-sequence statistics, column variance, and clustering consistency across models and behavioral groups.

\subsection{Comparison of Acoustic Feature Extractors and Developmental Groups}

Different feature extraction backbones handle the continuous, complex acoustic properties of dolphin vocalizations in distinct ways. Table \ref{tab:aggression_comparison} summarizes the performance metrics comparing the baseline small transformer model on adult aggression against alternative representations (MFCCs, Wav2vec \cite{wav2vec2}, Perch \cite{perch}) and the juvenile play aggression dataset, along with their respective relative changes. The results are shown in Table \ref{tab:aggression_comparison}.

\begin{table*}[t]
\centering
\caption{Comprehensive MSA quality comparison across feature extractors, model configurations, and developmental datasets. Values and relative changes (lifts \%) are derived from experimental alignment runs.}
\label{tab:aggression_comparison}
\begin{tabular}{l|c|cc|cc|cc|cc}
\hline
\textbf{Metric} & \textbf{Agg. Small} & \textbf{MFCC} & \textbf{Delta (\%)} & \textbf{Wav2vec} & \textbf{Delta (\%)} & \textbf{Perch} & \textbf{Delta (\%)} & \textbf{Juv. Small} & \textbf{Delta (\%)} \\
\hline
gap\_ratio & 0.268 & 0.0005 & -99.82 & 0.0006 & -99.78 & 0.0151 & -94.37 & 0.564 & +110.77 \\
gap\_opens & 379.5 & 104.0 & -72.60 & 102.0 & -73.12 & 310.0 & -18.31 & 310.8 & -18.10 \\
gap\_extends & 1592.0 & 221341 & +13803.3 & 176684 & +10998.2 & 31173.3 & +1858.1 & 1025.0 & -35.62 \\
total\_gaps & 1971.5 & 221445 & +11132.3 & 176786 & +8867.1 & 31483.3 & +1496.9 & 1335.8 & -32.24 \\
gap\_percentage (\%) & 55.72 & 97.22 & +74.49 & 97.37 & +74.75 & 81.68 & +46.59 & 41.18 & -26.09 \\
mean\_variance & $2.35 \times 10^{-5}$ & 0.0 & -100.0 & 0.0 & -100.0 & $1.13 \times 10^{-7}$ & -99.52 & $2.13 \times 10^{-5}$ & -9.41 \\
median\_variance & $2.06 \times 10^{-5}$ & 0.0 & -100.0 & 0.0 & -100.0 & $7.90 \times 10^{-8}$ & -99.62 & $1.57 \times 10^{-5}$ & -23.83 \\
percentage\_clean (\%) & 78.93 & 0.0 & -100.0 & 0.0 & -100.0 & 3.71 & -95.29 & 78.24 & -0.87 \\
total\_columns & 188.0 & 6327.0 & +3265.4 & 4778.0 & +2441.5 & 1509.0 & +702.7 & 102.8 & -45.32 \\
alignment\_rate (\%) & 68.97 & 62.07 & -10.0 & 65.52 & -5.0 & 67.24 & -2.5 & 80.21 & +16.30 \\
cluster\_mean\_purity & 0.757 & 1.0 & +32.16 & 1.0 & +32.16 & 0.995 & +31.43 & 0.859 & +13.51 \\
cluster\_pct\_pure\_cols (\%) & 42.59 & 100.0 & +134.8 & 100.0 & +134.8 & 98.68 & +131.7 & 61.11 & +43.50 \\
\hline
\end{tabular}
\end{table*}

\subsection{Model Comparison on Aggression Recordings}

The choice of feature extractor fundamentally dictates how sequence alignment algorithms interpret aggressive vocalizations, particularly the characteristic synchronized burst pulses. The whisper small model achieves a balanced gap ratio (0.268) and a moderate gap percentage (55.72\%), maintaining a high percentage of clean columns (78.93\%) which ensures that biological structures remain interpretable, without over-fragmenting the timeline, on the adult aggression dataset. In contrast, traditional spectral representations like MFCCs and self-supervised speech models like Wav2vec yield extreme structural shifts when forced into the alignment framework; they exhibit massive increases in gap extensions (surpassing 170,000 to 220,000 extensions) and drive the overall gap percentage above 97\%. While this forces a rigid, mathematically pure cluster assignment (achieving 100\% cluster purity and 0 column variance due to excessive stretching and padding), it severely dilutes temporal resolution by expanding the total alignment length to thousands of sparse columns. Meanwhile, the bioacoustic foundation model Perch strikes a middle ground: it significantly reduces the gap ratio down to 0.015 (-94.37\%) and elevates cluster purity to 99.45\%, though it still incurs heavier gap insertion loads than the fine-tuned transformer encoder. Overall, while general-purpose or raw feature models optimize mathematical homogeneity, the domain-adapted transformer encoder provides the most structurally balanced representation for uncovering natural temporal motifs in aggressive dolphin interactions.

\subsection{Comparison with Juvenile Play Aggression Data}
\label{sec:biology}
Comparing adult aggression profiles against the juvenile play aggression dataset reveals significant developmental differences in vocal control and structural consistency. As shown in Table~\ref{tab:aggression_comparison}, the juvenile  dataset exhibits a substantially higher gap ratio (0.564, a +110.77\% increase over adult aggression) alongside a reduced total column count (102.8 columns vs. 188.0). Despite having shorter overall sequence lengths, juvenile calls demonstrate higher alignment rates (80.21\% compared to 68.97\% for adults) and elevated cluster purity (0.859 mean purity versus 0.757). This pattern indicates that while young dolphins produce acoustically variable and less rigidly stereotyped signals—reflected in greater local gapping complexity—their overall repertoire clusters into broader, more inclusive categories. This result supports the developmental hypothesis that juvenile dolphins are still refining fine motor control over their vocal apparatus, resulting in higher intra-class variance compared to the highly crystallized, stereotyped call profiles of mature adults.

\section{Discussion}
\begin{figure*}[ht]
  \centering
  \includegraphics[width=\textwidth]{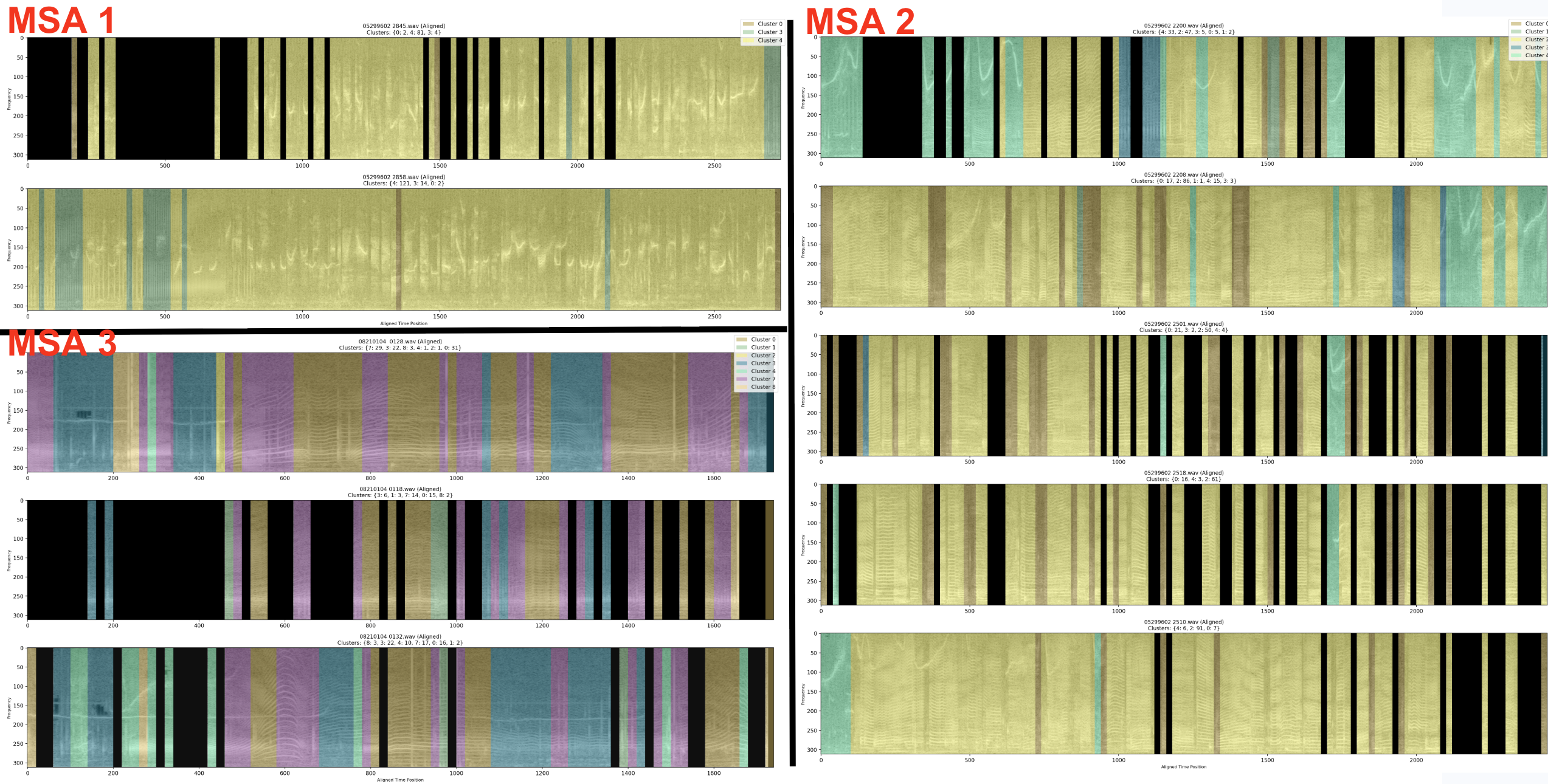}
  \caption{Multiple sequence alignment results for adult aggression vocalizations, producing three clusters (MSA 1, MSA 2, MSA 3). Each row is one aligned recording; the x-axis is alignment position (equivalent to time within each recording, in ms, after dynamic-programming alignment — not absolute recording time). Colored overlays mark the spectral cluster assigned to each acoustic frame by the column-informed spectral clustering (Section \ref{sec:clustering}): yellow = Cluster 0, cyan = Cluster 1, magenta = Cluster 2 (see legend inset). Solid black blocks denote gap positions inserted by the alignment algorithm to preserve temporal correspondence across recordings and carry no acoustic content. Vertically consistent color bands across rows indicate a conserved acoustic motif recovered at that alignment column, despite the motif occurring at different absolute times in the raw recordings. MSA 1 and MSA 2 show largely homogeneous (single-cluster, yellow) structure, consistent with stereotyped call types; MSA 3 shows a mix of clusters (cyan, yellow, magenta), indicating greater within-group acoustic variability)}
  \label{fig:alignment_results}
\end{figure*}

Our experiments demonstrate that multiple sequence alignment techniques, originally developed for discrete biological sequences, can be effectively adapted to analyze continuous dolphin vocalizations. The resulting visualizations reveal structural patterns that are difficult to detect through standard spectrogram inspection, providing marine mammalogists with a new tool for comparative acoustic analysis.

Figure \ref{fig:alignment_results} shows the output of our pipeline applied to aggressive vocalizations, producing three MSA clusters. MSA 1 and MSA 2 show homogeneous, stereotyped structure, while MSA 3 shows greater acoustic diversity. See the figure caption for details on cluster color coding and interpretation.

Our evaluation of different feature extraction backbones reveals that the choice of acoustic representation fundamentally dictates multiple sequence alignment quality. Traditional feature representations and self-supervised speech models—specifically MFCCs and Wav2vec—yield extreme structural shifts when integrated into the alignment framework, driving gap percentages above 97\% and vastly expanding the total alignment length to thousands of sparse columns. While this forces a rigid, mathematically pure cluster assignment with zero column variance, it severely dilutes temporal resolution and introduces excessive gap extensions. In contrast, the bioacoustic foundation model Perch and our fine-tuned domain-specific Aggression whisper small transformer strike a much more balanced trade-off. The whisper small transformer achieves a moderate gap percentage (55.72\%) and a high proportion of clean columns (78.93\%), preserving natural acoustic structures without over-fragmenting the timeline. Meanwhile, Perch significantly reduces the gap ratio down to 0.015 and elevates cluster purity to 99.45\%. Overall, these results demonstrate that while generic or raw feature representations optimize artificial mathematical homogeneity, domain-adapted transformer embeddings provide the most structurally robust representation for capturing natural temporal motifs in aggressive dolphin interactions.  

The comparison between adult and juvenile calls (Section \ref{sec:biology}) supports the hypothesis that vocal stereotypy increases with age, consistent with juvenile dolphins still acquiring motor control over their vocal apparatus.

Several limitations warrant consideration. First, our evaluation metrics assess alignment quality but do not directly measure biological meaningfulness; low gap percentages and consistent clustering do not guarantee that the identified correspondences reflect functionally homologous acoustic events. Second, the reliance on a transformer encoder pre-trained on human speech introduces potential biases toward acoustic features relevant for human language rather than dolphin communication. While fine-tuning with dolphin-specific data mitigates this concern, the optimal architecture for cetacean vocalization analysis remains an open question. Third, our early stopping criterion for cluster formation requires manual threshold selection. Finally, the current pipeline operates on pre-segmented recordings containing isolated vocalizations; extending the
approach to continuous recordings with multiple overlapping vocalizers presents additional challenges for source separation and sequence boundary detection.
Our listener-first framework deliberately constrains the applications of this research. By focusing on pattern recognition rather than signal synthesis, we avoid the ethical risks associated with playback experiments while still providing tools that enhance human understanding of dolphin social states. The identification of aggressive acoustic patterns, for instance, could help researchers recognize escalating tensions within a pod and adjust their observation protocols accordingly, reducing the risk of inadvertent interference. However, we emphasize that the patterns identified through MSA should inform passive observation rather than active intervention; the complexity of dolphin social dynamics precludes confident predictions about how individuals will respond to specific acoustic stimuli.

\section{Future Work}

Several promising directions extend the current framework. First, profile hidden Markov models (pHMMs) \cite{durbin} offer a natural extension of multiple sequence alignments, providing probabilistic representations of conserved motifs that can be used to classify novel recordings or detect specific call types in continuous audio streams. Training pHMMs on the alignments produced by our pipeline would enable automated recognition of behavioral states based on acoustic signatures.

Second, grammar induction techniques could leverage the sequential structure revealed by MSA to identify combinatorial rules governing vocalization production. If dolphin calls exhibit compositional structure, where complex utterances are assembled from simpler acoustic units according to consistent rules, alignment-based methods provide an appropriate foundation for discovering these patterns. The cluster labels assigned to each MSA column effectively discretize the continuous acoustic stream, producing symbolic sequences suitable for grammar induction algorithms originally developed for language acquisition research.

Third, extending the framework to multi-modal data would enable joint analysis of acoustic and behavioral observations. Aligning vocalizations with simultaneously recorded video could reveal correlations between specific acoustic motifs and physical behaviors, providing richer context for interpreting the communicative function of different call types.

Finally, comparative analysis across species would test whether the structural patterns observed in spotted dolphins generalize to other cetaceans. Applying the same alignment methodology to recordings from bottlenose dolphins, orcas, or other vocal learners could reveal both shared acoustic principles and species-specific adaptations, contributing to broader questions about the evolution of complex communication systems.

\section{Conclusion}

We have presented a framework for analyzing dolphin vocalizations through multiple sequence alignment, adapting the ClustalW algorithm to operate on continuous acoustic data represented as transformer-derived embeddings. By replacing discrete substitution scores with Gaussian kernel similarity measures, our approach generates visualizations that reveal shared temporal structure across recordings, highlighting conserved motifs that are difficult to detect through standard spectrogram inspection.

Our experiments demonstrate that the choice of feature representation fundamentally dictates multiple sequence alignment quality. Standard bioacoustic baselines and self-supervised models—such as MFCCs and Wav2vec—yield extreme structural expansion with high gap percentages, whereas domain-adapted transformer encoders and bioacoustic foundation models like Perch achieve a much more balanced structural representation. Furthermore, the comparison between adult aggression and juvenile play aggression calls provides quantitative evidence for developmental changes in vocal stereotypy, with juvenile play calls exhibiting distinct structural variability and higher gapping ratios compared to adult vocalizations. These comprehensive benchmarks validate multiple sequence alignment as a robust and versatile tool for comparative bioacoustics, offering marine mammalogists new methods for discovering structure in large vocalization databases across diverse feature configurations.  

By adhering to a listener-first ethical framework, our approach prioritizes improving human recognition of dolphin social states without risking the disruption that could result from playback experiments. The patterns revealed through MSA inform passive observation and non-invasive research, supporting conservation efforts while respecting dolphin autonomy and the integrity of wild behavioral repertoires. As acoustic monitoring technology continues to improve and vocalization databases grow, alignment-based methods offer a principled approach to extracting meaningful structure from the complexity of cetacean communication.
\bibliographystyle{unsrt}
\bibliography{bibliography}

\end{document}